\documentclass{article}

\usepackage{arxiv}

\usepackage[utf8]{inputenc} 
\usepackage[T1]{fontenc}    
\usepackage{hyperref}       
\usepackage{url}            
\usepackage{booktabs}       
\usepackage{amsfonts}       
\usepackage{nicefrac}       
\usepackage{microtype}      
\usepackage{lipsum}
\usepackage{graphicx}
\graphicspath{ {figures/} }

\usepackage{times}
\usepackage[square,numbers]{natbib}
\usepackage{amssymb} 
\usepackage{amsmath}
\usepackage{tikz}
\usetikzlibrary{arrows.meta} 
\usepackage[nomarkers,figuresonly, nofiglist]{endfloat}

\newcommand{\nTotal}{2000} 
\newcommand{\Totalmone}{1999}

\title{Assessing the Impact of Covariate Distribution and Positivity Violation on Weighting-Based Indirect Comparisons: a Simulation Study}

\author{
 Arnaud Serret-Larmande \\
Sorbonne University, INSERM,  \\
Pierre Louis Epidemiology and Public Health Institute \\
AP-HP, Pitié-Salpêtrière Hospital, Public Health Department \\
PSL-CFX Clinical Research Unit, CIC-1901 \\
Paris, France  \\
\texttt{arnaud.serretlarmande@gmail.com} \\
   \And
 Jérôme Lambert \\
  ECSTRRA Team IRSL \\
  INSERM UMR 1342, Université Paris Cité \\
  Hôpital Saint-Louis \\
  Paris, France  \\
  \texttt{jerome.lambert@aphp.fr}
  \And
 Stéphane Gaudry \\
 AP-HP, Hôpital Avicenne, \\
 Service de Réanimation Médico-Chirurgicale  \\
 INSERM, UMR-S 1155 CORAKID \\
 Paris, France \\
 \texttt{stephane.gaudry@aphp.fr}
\And
 David Hajage \\
Sorbonne University, INSERM, \\
Pierre Louis Epidemiology and Public Health Institute \\
AP-HP, Pitié-Salpêtrière Hospital, Public Health Department \\
PSL-CFX Clinical Research Unit, CIC-1901 \\
Paris, France  \\
\texttt{david.hajage@aphp.fr}
}
\begin{document}
\maketitle
\begin{abstract}
Population-Adjusted Indirect Comparisons (PAICs) are used to estimate treatment effects when direct comparisons are infeasible and individual patient data (IPD) are only available for one trial. Among PAIC methods, Matching-Adjusted Indirect Comparison (MAIC) is the most widely used. However, little is known about how MAIC performs under challenging conditions such as limited covariate overlap or markedly non-normal covariate distributions.

We conducted a Monte Carlo simulation study comparing three estimators: (i) MAIC matching first moment (MAIC-1), (ii) MAIC matching first and second moments (MAIC-2), and (iii) a benchmark method leveraging full IPD---Propensity Score Weighting (PSW). We examined eight scenarios ranging from ideal conditions to situations with positivity violations and non-normal (including bimodal) covariate distributions. We assessed both anchored and unanchored estimators and examined the impact of adjustment model misspecification. We also applied these estimators to real-world data from the AKIKI and AKIKI-2 trials, comparing renal replacement therapy strategies in critically ill patients.

MAIC-1 demonstrated robust performance, remaining unbiased in the presence of moderate positivity violations and non-normal covariates, while MAIC-2 and PSW appeared more sensitive to positivity violations. All methods showed substantial bias when key confounders were omitted, emphasizing the importance of correct model specification. In real-world data, a consistent trend was found with MAIC-1 showing narrower confidence intervals with positivity violation.

Our findings support the cautious use of unanchored MAICs and highlight MAIC-1’s resilience across moderate violations of assumptions. However, the method's limited flexibility underscores the need for careful use in real-world settings.
\end{abstract}


\section{Introduction}\label{sec1}

When estimating the relative effect of two treatments of interest, the gold standard is typically a randomized controlled trial (RCT). However, drawing evidence from an RCT is not always feasible due to various constraints, including circumstances that justify the use of a single-arm trial, proliferation of available treatment options, as well as the need to generate evidence rapidly in contexts of emerging therapies. Indirect Treatment Comparisons can be used to estimate relative effects of treatments that have been evaluated across different studies. They are used heavily in the context of Health Technology Assessment as substitutes when direct evidence is unavailable 
\citep{macabeoAcceptanceIndirectTreatment2024}
\citep{igarashiIndirectTreatmentComparisons2025} \citep{vanierRapidAccessInnovative2023}. Since such comparisons rely on outcomes observed across distinct populations, they are inherently subject to confounding bias. One approach to mitigate this bias is to use a common treatment arm evaluated across two RCTs (Bucher Indirect Treatment Comparisons) \citep{bucherResultsDirectIndirect1997}. This shared treatment arm is referred to as an anchor, and the resulting analyses are known as anchored comparisons. While this method cancels out the bias related to differences in the distribution of prognostic factors across trials, the estimated treatment effect remains biased in the presence of treatment-effect modifiers. Moreover, it requires by definition the presence of a common treatment arm. Consequently, an anchored comparison is impossible when the studies lack a common treatment arm---which will always be the case when a single-arm study is used in the comparison, an increasingly frequent situation \citep{beaver25YearExperienceUS2018} \citep{agrawalUseSingleArmTrials2022} \citep{luConsiderationsSingleArmTrials2024}.

Adjusted estimators are thus necessary to control for confounding and obtain an unbiased treatment effect. Such adjustment can be carried out using a balancing score (\textit{e.g.} propensity score) or outcome regression models. Usually, these estimators would require having access to complete Individual Patient Data (IPD).

However, in the context of Health Technology Assessment, sponsors have access to IPD for the treatment that they are developing, but often they can only access published aggregated data for treatments developed by competitors. This explains the increasing use of Population-Adjusted Indirect Comparisons (PAIC), which are methodologies designed to estimate adjusted indirect treatment effects in pairwise comparisons when IPD is not available for all the studies informing the comparison. Two main classes of estimators are used in this context: Simulated Treatment Comparison and Matching-Adjusted Indirect Comparison (MAIC). Simulated Treatment Comparisons are outcome regression-based models estimating average conditional treatment effect estimates \citep{caroNoHeadtoHeadTrial2010}, while MAICs use a propensity-score-like reweighting approach to estimate marginal effects \citep{signorovitchjComparativeEffectivenessResearch2012}. More recent methodological advancements such as Multilevel Network Meta-Regression allow estimating treatment effects in different target populations \citep{phillippoMultilevelNetworkMetaregression2024}. A large body of methodological literature has explored the properties and limitations of PAICs \citep{phillippoMethodsPopulationAdjustedIndirect2016} \citep{phillippoMethodsPopulationAdjustedIndirect2018} \citep{remiro-azocarMethodsPopulationAdjustment2021}, and highlighted their growing practical application in that context (\citep{truongPopulationAdjustedIndirect2023}, \citep{serret-larmandeMethodologicalReviewPopulationadjusted2023} \citep{farinassoMatchingAdjustedIndirectComparison2025}). Thanks to their abilities to adjust for confounding, PAICs can be extended to situations where a common treatment arm is missing. While estimates yielded by the anchored form are deemed much more reliable \citep{phillippoMethodsPopulationAdjustedIndirect2016}, unanchored PAICs have been particularly popular due to their theoretical ability to derive treatment effect estimates even in situations where a common comparator arm is missing.

This work focuses exclusively on MAIC, as it remains the most widely used PAIC estimator in the literature and in applied settings \citep{serret-larmandeMethodologicalReviewPopulationadjusted2023}, \citep{truongPopulationAdjustedIndirect2023}. Briefly, MAIC estimates balancing weights using the method of moments and then reweights the Individual Patient Data (IPD) sample in a way that ensures that the weighted covariate moments (typically first, and sometimes second-order moment) match those of the Aggregated Data (AgD) trial. Relative treatment effect is subsequently estimated by comparing the weighted outcome from the IPD trial to the observed outcome in the AgD trial.

Nonetheless, despite MAICs appealing simplicity, the requirement to respect usual causal inference assumptions remains unchanged, that is exchangeability, consistency, and positivity \citep{hernanCausalInferenceWhat2020}. Moreover, the lack of fully accessible individual data imposes important limitations. First, the target estimand is restricted by design to be the treatment effect in the AgD population. Second, the estimator itself may be constrained by the moments-matching approach (more details in the Methods section). Theoretical properties of estimators based on the method of moments may not hold when covariate distributions are bimodal or with heavy-tail distributions \citep{yuHarmonicTransformbasedDensity2023}
 \citep{asquithLmomentsTLmomentsGeneralized2007}. Additionally, the logistic regression model used for weight estimation is fit on a subset of the data of interest (IPD trial), limiting statistical efficiency.

These constraints raise important concerns about the robustness of MAIC when its theoretical assumptions are violated. In such settings, a natural question arises: would access to full IPD from both trials yield a more reliable estimate as compared to partially AgD?

To date, however, no study has been conducted to compare MAICs to its full-IPD counterpart, \textit{i.e.} Propensity Score Weighting (PSW) with an estimand targeting the AgD trial population, that is using average treatment effect in the controls weights \citep{austinIntroductionPropensityScore2011}.

This study addresses this gap by comparing the performance of MAIC and PSW across a range of scenarios using Monte Carlo simulations. Additionaly, the relevance of these methods are assessed using empirical data from the AKIKI and AKIKI-2 clinical trials, which compared strategies for initiating renal replacement therapy in critically ill patients with acute kidney injury.

\section{Methods}
\subsection{Notation and Identification}

The goal is to estimate the relative marginal effect of treatment $A$ versus treatment $B$, noted $\Delta$, as measured by an outcome $Y$.

These treatments have been evaluated in two separate clinical trials $T \in \{a, b\}$. Both $a$ and $b$ trials also evaluated a common treatment arm $C$. The treatment received is noted ${Z}$, with $Z_a \in \{A, C\}$, and $Z_b \in \{B, C\}$ in trials $a$ and $b$ respectively; as an example, an observation of $Y$ under a treatment $Z=A$ in trial $T=a$ will be written $y_a(Z=A)$, simplified as $y_a(A)$ thereafter.

The population of interest is the source population of the $b$ clinical trial. Therefore the target estimand for the indirect comparison is the treatment effect in the $b$ population: $\Delta_{b}$. It is known in the propensity score literature as the Average Treatment effect in the Control population (ATC) \citep{zengMovingBestPractice2025}. It entails specific implications for the positivity assumption by introducing an asymmetry in the required support for the baseline characteristics one wishes to balance across studies: under this weighting scheme, the positivity assumption holds as long as the covariate distribution in the $b$ target trial is fully represented within the $a$ index trial. Conversely the $a$ trial covariate distribution may not need to be represented in the $b$ trial. This contrasts with the Average Treatment Effect, which requires complete overlap in the covariate distribution between both trials.

In the unanchored case, an unbiased estimate of $\Delta_{b} $ is given by $\hat\Delta_{b} = g(\bar y_{b}(A)) - g(\bar y_{b}(B))$, with $\bar y_b(A)$ and $\bar y_b(B)$ being the sample mean of $Y$ under respectively treatment $A$ and $B$ in the trial $b$, and $g$ being an invertible  link function that ensures additivity of outcomes.
For instance, if $Y$ is normally distributed, $g$ would be the identity function ($g(x) = x$), and if $Y$ is binary, $g$ could be the identity function (estimating a difference of proportions), log function ($g(x) = log(x)$) when estimating a log relative risk), or logit function ($g(x) = log(\frac{x}{1-x})$) when estimating a log odds ratio. Because these Monte Carlo simulations' target estimand is the mean difference of continuous variables, we use the identity link and drop $g(\cdot)$ in what follows.

The estimation problem arises from the fact that the quantity $\bar y_{b}(A)$ is unobserved, that is treatment $A$ has not been evaluated in trial $b$. Using the sample mean of $Y$ under treatment $A$ observed in $a$ trial $a$, \textit{i.e. }$ \bar y_a(A)$, as a substitute for $\bar y_b(A)$ will yield a biased estimate of $\Delta_b$ if there exist prognostic factors (PF) in imbalance between $a$ and $b$ studies, which is most certainly the case when considering two separate clinical trials. Some of these PF may additionally influence the outcome differently depending on the treatment received, defining treatment-effect modifiers (TEM). Therefore, in practice, an unbiased estimation of $\Delta_{b}$ requires to estimate $\hat{\bar y}_b(A)$, that is to get an estimation of the average value of the outcome that would have been observed under treatment $A$ in the $b$ trial, \textit{i.e.} in a sample with distributions of PF and TEM representative of the $b$ source population. $\hat{\bar y}_b(A)$ is unbiased (and subsequently so is $\hat \Delta_b$) if the estimation process adequately takes into account the distribution of the PF and the TEM (that is all the confounders of the association between the treatment and the outcome) in the $b$ trial population.

This requirement can be relaxed when using the common treatment arm $C$, which plays the role of an anchor. In such a case, the estimation of $\Delta_b$ can be obtained through a difference-in-difference between the outcomes observed under each treatment across the two trials $a$ and $b$, that is $\hat \Delta_{b}  = \left(\bar y_b(A) - \bar y_b(C) \right) -  \left(\bar y_b(B) - \bar y_b(C) \right)$. The unobserved quantity $\left( \bar y_b(A) - \bar y_b(C) \right)$, that is the relative difference of the outcome $Y$ under treatments $A$ versus $C$ in the $b$ population, can be substituted by $\left(\hat{\bar y}_b(A) - \hat{\bar y}_b(C) \right)$, that is its estimation would have treatment $A$ had been evaluated in trial $b$. The quantity $\left(\hat{\bar y}_b(A) - \hat{\bar y}_b(C) \right)$ is an unbiased estimate of $\left(\bar y_b(A) - \bar y_b(C)\right)$ (and subsequently $\hat \Delta_b$ of $\Delta_b$) if the estimation process adequately takes into account the distribution of the TEM---and not necessarily of the PF, in the $b$ trial population.


\subsection{Estimators}

Three different treatment effect estimators will be used to estimate $\Delta_b$.
We define the unweighted estimator, the MAIC estimator, and the PSW estimator targeting the marginal effect in the trial $b$ population.
The goal of each estimator is to provide an estimate of $\Delta_b$ despite the lack of direct observations of $\bar y_b(A)$. Each estimator is assessed under both anchored and unanchored specifications.

\subsubsection{Unweighted Estimator}
The unweighted approach ignores cross-trial population differences and uses trial $a$’s observed outcome under a given treatment as a proxy for the outcome that would have been observed in trial $b$ under that given treatment. This yields the unanchored case: $\hat \Delta_{b} = \bar y_{a}(A) - \bar y_{b}(B)$, and the anchored case: $\hat \Delta_{b}  = \left( \bar y_a(A) - \bar y_a(C) \right) -  \left(\bar y_b(B) - \bar y_b(C) \right)$.

\subsubsection{PSW Estimator}

PSW reweights each subject $i$ from trial $a$ with a weight $w_i$ so that the distribution of their baseline characteristics mirrors that of trial $b$. Each weight $w_i$ is equal to the odds that individual $i$ be included in trial $b$ rather than in trial $a$, given their covariates. Individuals from trial $b$ are assigned a weight of 1. If $e(x)=\Pr(T=b \mid X =x)$ is the probability of being in trial $b$, then for an individual with a vector of covariates $X_i$,

\begin{equation}
    w_i = \frac{Pr(T_i= b \mid X_i)}{Pr(T_i = a \mid X_i)} = \frac{ e(X_i)}{1 -  e(X_i)}
\end{equation}

In practice, the estimate of $\frac{e(X_i)}{1 - e(X_i)}$ corresponds to the exponential of the linear predictor from a logistic regression model including covariates $X_i$, which models the probability of being included in trial $b$. The model is fitted by maximum likelihood using IPD from both trials.

Finally, we estimate the outcome under treatment $z$ in the $b$ trial population with:

\begin{equation}
    \hat{\bar y}_b(z) = \frac{\sum_{i}  w_{i} y_{i} \mathbb{I}(Z_i=z)}{{\sum_{i}  w_{i} \mathbb{I}(Z_i=z)}}
    \label{eq:treatment_effect_ps}
\end{equation}

The treatment effect $\hat \Delta_b$ is then estimated by replacing the estimated quantities from the $b$ trial with the estimated ones, as detailed in the Notations and Identification section.

\subsubsection{MAIC Estimators}

MAIC estimator targets the same estimand as PSW, that is the Average Treatment effect in the Control population. However, unlike PSW, it relies on moment-matching rather than full-distribution balancing to yield this treatment effect estimate. Each subject $i$ from trial $a$ is given a weight $w_i$ such that the moments of the weighted distribution of baseline covariates $X$ in trial $a$ matches the reported baseline distribution moments in trial $b$. Formally, $w_i$ is defined to be the odds that individual $i$ be included in trial $b$ rather than in trial $a$, given $X_i$. Keeping $e(x)$ definition we get:

\begin{equation}
    \tilde w_i = \frac{Pr(T_i= b \mid h(X_i))}{Pr(T_i = a \mid h(X_i)} = \frac{e(X_i)}{1 - e(X_i)}
\end{equation}

 with $h$ a function yielding the vector of moments for a vector of covariates $X$ \citep{signorovitchComparativeEffectivenessHeadtohead2010}. In practice, $\tilde w_i$ is obtained by solving method of moments equations that equate weighted covariate moments in trial $a$ to the covariate moments reported for trial $b$. Let $h(X_b)$ be the set of relevant baseline summary measures available in trial $b$ (e.g. means and eventually higher moments), MAIC estimates $\tilde {w}_i$ such that the weighted vector of baseline covariates $X_a$ in trial $a$ equates the vector of moments of baseline covariates $X_b$ in trial $b$:

\begin{equation}
    h(\tilde w_i X_{i, a}) = h(X_b)
\end{equation}

In practice, this is done by finding the value of the vector $\alpha$ that minimizes the following sum: $\sum_i^n exp\left(\alpha^T \left(X_i - h(X_b) \right) \right)$.
Two configurations were tested for the vector of moments $h(X_b)$: (i) first-order moments only and (ii) both first- and second-order moments. Including only the first moments corresponds, in practice, to matching covariate means, while including both the first and second moments corresponds to matching both means and variances. We consider these two settings to represent two different estimators, and we will refer to them as respectively MAIC-1 and MAIC-2 thereafter.

The estimated outcome under treatment $Z=z$ is then formed using Equation \ref{eq:treatment_effect_ps}, and relative treatment effect $\Delta_b$ is estimated by substituting unobserved with estimated quantities.

\subsubsection{Variance of the treatment effect}
The computation of the variance of the treatment effect is identical for the three different estimators and is the sum of the variances of the outcome under each treatment used in the indirect comparison. Variance of the outcome in trial $b$ under treatment $z$ (weighted or unweighted) has been estimated with 2,000 non-parametric bootstrap iterations.


\subsection{Data Generation Mechanisms (DGMs)}

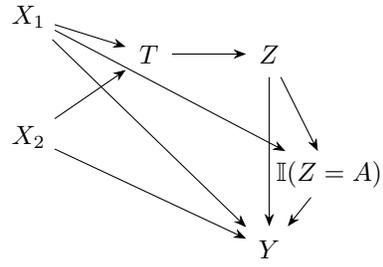
\begin{figure}[ht]
\centering
\begin{tikzpicture}[%
    ->,              
    >=Stealth,       
    node distance=1.6cm,  
    every node/.style={draw=none, minimum size=6mm} 
]

\node (X1) {$X_1$};
\node (X2) [below of=X1] {$X_2$};
\node (T)  [right of=X1, yshift=-0.5cm] {$T$};
\node (Z)  [right of=T] {$Z$};
\node (Y)  [below of=Z, yshift=-1.0cm] {$Y$};
\node (I) [below of=Z, xshift=0.8cm] {$\mathbb{I}(Z=A)$};

\draw (X1) -- (T);
\draw (X2) -- (T);
\draw (X1) -- (Y);
\draw (X2) -- (Y);
\draw (T)  -- (Z);
\draw (Z)  -- (Y);
\draw (Z) -- (I);
\draw (X1) -- (I);
\draw (I) -- (Y);

\end{tikzpicture}
    \caption{Relationships among \(X_1, X_2, T, Z,\) and \(Y\).}
\label{fig:dag}
\end{figure}

Following the previously introduced notation, Figure~\ref{fig:dag} depicts the causal Directed Acyclic Graph (DAG) of the true relationships between the variables used in our simulations: baseline covariates $X_1$ and $X_2$, trial assignment $T$, treatment $Z$, and outcome $Y$.

\begin{figure}
    \centering
    \includegraphics[width=1\linewidth]{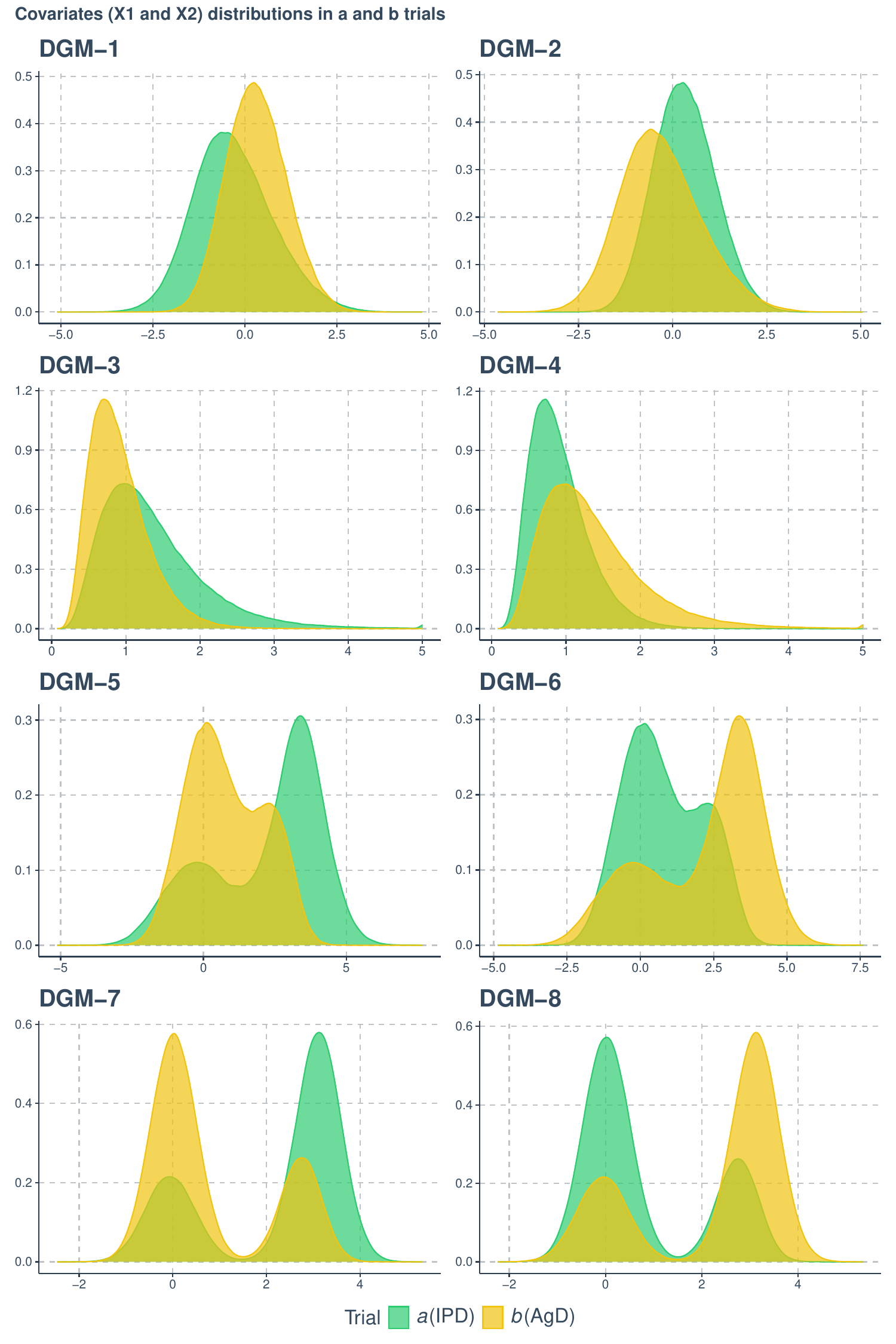}
    \caption{Covariate distributions depending on DGM}
    \label{fig:covariates_distributions}
\end{figure}

From this DAG, we defined eight different DGMs that differ by the distributions of the covariates $X_1$ and $X_2$ and by the $b$ trial assignment model. In every DGMs, $X_1$ is a PF and a TEM, and $X_2$ is a PF only---it does not influence the treatment effect. In every scenario, the outcome model will be:
\[
Y = X_1 + 2 X_2 + (1 + 2 X_1) I(Z=A) + I(Z=B) + \epsilon
\]

with $\epsilon \sim N(0,1)$. The treatment $Z$ is randomly allocated in a 1:1 ratio within each trial, with the two treatments arms being $A$ and $C$ in the $a$ trial---\textit{i.e.} $\mathbb{P}(Z = A |T=a) = \mathbb{P}(Z = C|T=a) = 0.5$, and $B$ and $C$ in the $b$ trial---\textit{i.e.} $\mathbb{P}(Z = B|T=b) = \mathbb{P}(Z = C|T=b) = 0.5$.
Within each DGM, $X_1$ and $X_2$ are independent and identically distributed. The eight scenarios examined can be grouped into four pairs. Each pair is distinguished by $X_1$ and $X_2$ distributions and their effect on the trial assignment probability. Within each pair, one scenario presents positivity issues and the other not, as detailed below. Consistent with the explanation given in the Notation and Identification section, violation of the positivity assumption is defined as a lack of statistical support in trial $a$ for the covariate distribution observed in trial $b$---which represents the target population. Below, we summarize the characteristics of the different situations explored by each scenario; additionally, the actual distribution of the covariate in trials $a$ and $b$ for each DGM is depicted in Figure \ref{fig:covariates_distributions}.

\noindent\textbf{DGM-1 and DGM-2 specifications}

The first (DGM-1) and second (DGM-2) scenarios feature two covariates $X_1$ and $X_2$ with standard normal distributions. DGM-1 serves as a control case with no violation of positivity, meaning complete support of the target trial $b$ in the source trial $a$. Conversely, DGM-2 introduces positivity violations due to a lack of statistical support, meaning part of $X_1$ and $X_2$ distributions in trial $b$ are not represented within the trial $a$. In the following equations, the parameter $\delta$ controls the direction of the trial assignment mechanism. A value of $\delta=-1$ reflects scenarios with limited overlap across trials, mimicking positivity violations. $\delta =1$ in DGM-1, and $\delta = -1$ in DGM-2.  

\begin{equation}
\begin{cases}
X_j &\sim N(0, 1), \quad j=\{1,2\} \\
T &\sim Bernoulli(\pi_T) \text{ where } \pi_T = P(T = b \mid X_1, X_2) = g \left( \delta \left[ \beta_0 + X_1 - 0.5 X_1^2 + X_2 - 0.5 X_2 \right] \right)
\end{cases}
\end{equation}

For all the DGMs explored, $g(x) = \left(1 + exp(-x) \right)^{-1}$, and the constant $\beta_0$ is calibrated so that the marginal probability of being assigned to trial $b$ is 0.5 in the overall population. Its numerical value is obtained 
using a root-finding algorithm.

\noindent\textbf{DGM-3 and DGM-4 specifications}
To explore the impact of non-normally distributed covariate, scenarios DGM-3 and DGM-4 use right-truncated log-normal distributions. In these scenarios, $\delta=-1$ in DGM-3, and $\delta = 1$ in DGM-4, that is DGM-4 presents positivity issues.

\begin{equation}
\begin{cases}
X_j &= min(X'_j, 5) \text{ with } X'_j \sim LogNormal(0, 0.5), \quad j=\{1,2\} \\
T &\sim Bernoulli(\pi_T) \text{ where } \pi_T = P(T = b \mid X_1, X_2) = g \left( \delta \left[2 \beta_0 + 2 X_1 + 2 X_2 \right] \right) \end{cases}
\end{equation}

\noindent\textbf{DGM-5, DGM-6, DGM-7 and DGM-8 specifications}
To examine the impact of bimodal covariate, scenarios five to eight generate two covariates with bimodal distributions. Scenarios 7 and 8 present more extreme bimodal distributions as compared to the fifth and sixth. DGM-5 and DGM-6 follow distributions laid out in Equation \ref{eq:DGM_5_6}, and DGM-7 and DGM-8 follow the ones in Equation \ref{eq:DGM_7_8}. $\delta=1$ in DGM-5 and DGM-7, and $\delta = -1$ in DGM-6 and DGM-8, that is DGM-6 and DGM-8 present positivity issues.

\vspace{-0.5em}
\begin{equation}
\begin{cases}
X_j &\sim \begin{cases}
        \mathcal{N}(0, 0.5) & \text{with probability } 0.5 \\
        \mathcal{N}(3, 0.5) & \text{with probability } 0.5 \\
        \end{cases}
        \quad j=\{1,2\} \\
T &\sim Bernoulli(\pi_T) \text{ where } \pi_T = P(T = b \mid X_1, X_2) = g \left( \delta \left[0.5 \beta_0 + 0.5 X_1 -  0.5 X_1^2 + 0.5 X_2 - 0.5 X_2^2 \right] \right) \\
\end{cases}
\label{eq:DGM_5_6}
\end{equation}

\begin{equation}
\begin{cases}

X_j &\sim \begin{cases}
        \mathcal{N}(0, 1) & \text{with probability } 0.5 \\
        \mathcal{N}(3, 1) & \text{with probability } 0.5 \\
        \end{cases}
        \quad j=\{1,2\} \\
T &\sim Bernoulli(\pi_T) \text{ where } \pi_T = P(T = b \mid X_1, X_2) = g \left( \delta \left[0.5 \beta_0 + 0.5 X_1 -  0.5 X_1^2 + 0.5 X_2 - 0.5 X_2^2 \right] \right) \\
\end{cases}
\label{eq:DGM_7_8}
\end{equation}

\subsubsection{Summary of scenarios}

A summary of the different DGMs with regards to covariate distribution, positivity issues and theoretical treatment effect value is given in Table \ref{tab:dgms-deltab}. The theoretical $\Delta_b$ value is the marginal treatment effect calculated using the real outcome model in a large population of $2\times 10^6$ subjects.

\begin{table}[ht]
\centering
\begin{tabular}{|c|c|c|c|}
\hline
 & \textbf{Covariate distribution} & \textbf{Positivity violation?} & \textbf{Theoretical treatment effect $\Delta_b$} \\
\hline
DGM-1 & Normal distribution    & No  & $\Delta_b = 0.68$ \\
DGM-2 & Normal distribution    & Yes & $\Delta_b = -0.55$ \\
DGM-3 & Lognormal distribution & No  & $\Delta_b = 1.17$ \\
DGM-4 & Lognormal distribution & Yes & $\Delta_b = 3.53$ \\
DGM-5 & Bimodal distribution   & No  & $\Delta_b = 1.52$ \\
DGM-6 & Bimodal distribution   & Yes & $\Delta_b = 2.27$ \\
DGM-7 & Bimodal distribution   & No  & $\Delta_b = 0.1$ \\
DGM-8 & Bimodal distribution   & Yes & $\Delta_b = 3.52$ \\
\hline
\end{tabular}
\caption{Summary of the different data-generating mechanisms (DGMs) and corresponding theoretical treatment effect}
\label{tab:dgms-deltab}
\end{table}

\subsection{Monte Carlo simulations settings} 

For each of the eight specified DGMs, 2,000 Monte Carlo iterations were performed. In each iteration, two trials of 250 patients per arm have been sampled from a super-population of $2 \times 10^{6}$ subjects, with a 1:1 randomization ratio (\textit{i.e.} 500 patients per trial in the unanchored setting and 1,000 per trial in the anchored one).

Additionally, in each setting, three sets of variables were used to adjust for: $\{X_1\}$, $\{X_2\}$, or $\{X_1; X_2\}$. These different sets correspond, respectively, to adjusting for treatment effect modifiers, prognostic factors, or both. Consequently, and depending on the treatment effect estimator used, this results in scenarios where the trial-assignment model is either correctly specified or misspecified.

\subsection{Performance metrics}

The objective of the simulation study is to evaluate the bias and precision of each estimator of $\Delta_b$
. The estimator's performance in this regard will be evaluated using the following metrics:

\subsubsection{Bias}
\begin{equation}
\label{eq:bias}
\text{Bias} = \sum_{j=1}^{\nTotal} \frac{\hat{\Delta}_{b,j} - \Delta_b}{\nTotal}
\end{equation}

The bias of an estimator is the difference between the expected value of $\hat \Delta_b$ and the true value of the parameter it is estimating $\Delta_b$.

\subsubsection{Root Mean Square Error (RMSE)}
\begin{equation}
\label{eq:rmse}
RMSE = \sqrt{ \frac{1}{\nTotal} \sum_{j=1}^{\nTotal} (\hat{\Delta}_{b,j} - \Delta_b)^2 }
\end{equation}

The Root Mean Square Error (RMSE) is the square root of the average of the squared differences between the predicted and true values , and reflects both the bias and variability of the estimator.

\subsubsection{Variability Ratio (VR)}
\begin{equation}
\label{eq:vr}
VR = \frac{\sum_{j=1}^{\nTotal} \frac{\sigma_{\Delta_{b, j}}}{\nTotal}}{\sqrt{\sum_{j=1}^{\nTotal} \frac{(\hat{\Delta}_{b,j} - \Delta_b)^2}{\Totalmone}}}
\end{equation}

The variability ratio compares the average model-based estimation of the standard error $\sigma_{\Delta_{b}}$ to the empirical standard error, indicating potential under- or overestimation of uncertainty.

\subsubsection{95\% coverage rate ($CR_{95\%}$)}
\begin{equation}
\label{eq:cr}
CR_{95\%} = \frac{1}{\nTotal} \sum_{i=1}^{\nTotal}
    \mathbb{I} \left(
        \Delta_b \,\in\, [\hat{\Delta}_{b,j} - 1.96 \hat \sigma_{\Delta_{b, j}}
               \;,\;
               \hat{\Delta}_{b,j} + 1.96 \hat \sigma_{\Delta_{b, j}}]
     \right)
\end{equation}

The coverage rate is the proportion of confidence intervals that contain the true parameter value (and should be equal to 0.95 in the case of the 95\% coverage rate), combining adequate estimation of the parameter and adequate estimation of uncertainty.

\subsubsection{Post-weighting covariate distribution overlap}

A graphical comparison of the unweighted and weighted covariate distributions between trials $a$ and $b$ was conducted to assess how the different estimators affected covariate balance, particularly with respect to the positivity assumption.

\subsubsection{Software}

The simulations were developed using the R programming language, version 4.5.1.

\section{Results}

\subsection{Treatment effect estimations under different DGMs, assuming correct model specification}

Results presented in this section use the correct trial assignment model---that is anchored estimators are adjusting on treatment-effect modifiers only ($X_1$), and unanchored estimators are adjusting on both prognostic factors and treatment-effect modifiers ($X_1$ and $X_2$).

\begin{figure}
    \centering
    \includegraphics[width=1\linewidth]{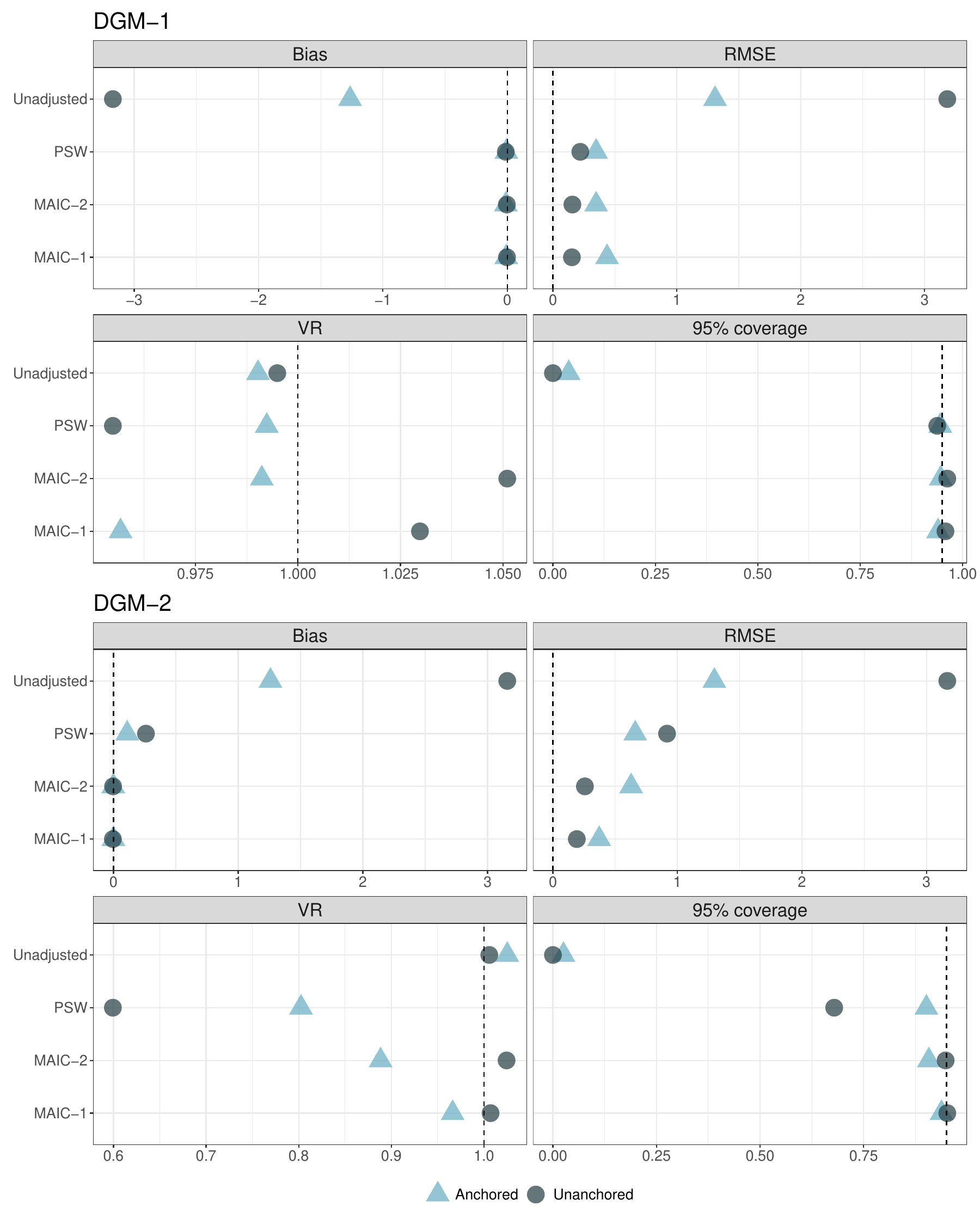}
    \caption{Treatment effect estimation under DGM-1 and DGM-2}
    \label{fig:treatment_scenario_1_2}
\end{figure}

Under DGM-1 (no overlap issue and covariate normally distributed), all adjusted estimators were unbiased (Figure \ref{fig:treatment_scenario_1_2}). As expected, estimates yielded by unanchored unweighted estimators were significantly more biased than the ones from anchored unweighted ones. All adjusted estimators exhibited close to nominal coverage rates, indicating well-calibrated confidence intervals. Finally, in this situation, unanchored showed lower RMSE, that is, more precise estimates. VR appeared closer to 1 with anchored as compared to unanchored estimators, hinting at a better variance estimation, even though the VR variation was small across the different estimators.

Under DGM-2, the target population includes regions of the covariate space that are poorly supported in source trial $a$, leading to violation of the positivity assumption (Figure \ref{fig:treatment_scenario_1_2}). In this situation, MAIC-1 and MAIC-2 produced unbiased estimates of the treatment effect, indicating robustness to moderate positivity assumption violations. RMSE was smaller for unanchored as compared to anchored MAICs. In contrast, PSW remained biased despite using the correct trial model, indicating larger sensitivity to violations of positivity assumptions. In such a case, RMSE was smaller with anchored as compared to unanchored. This suggests that anchoring on a common comparator arm helps mitigate the effects of limited common support, since it requires adjusting for TEM only (\textit{i.e.} only $X_1$). Regarding variance and coverage, nominal coverage rate was achieved with both unanchored MAIC estimators, and slight undercoverage with the anchored versions. PSW, affected by residual bias, failed to achieve acceptable coverage, especially with the unanchored version.

\begin{figure}
    \centering
    \includegraphics[width=1\linewidth]{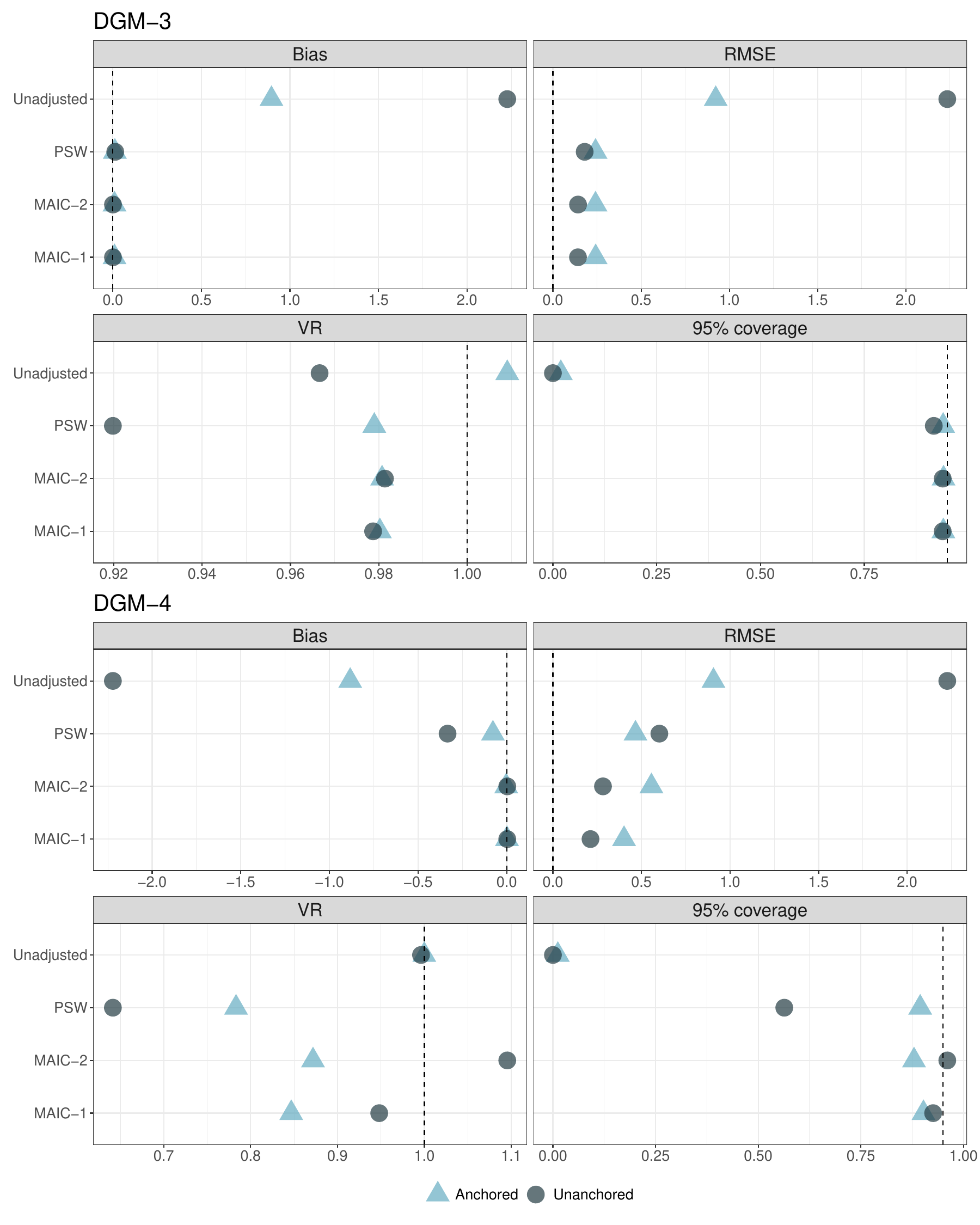}
    \caption{Treatment effect estimation under DGM-3 and DGM-4}
    \label{fig:treatment_scenario_3_4}
\end{figure}

Under DGM-3, which featured a lognormal distribution without positivity violations, estimators' performance was similar across all metrics (Figure \ref{fig:treatment_scenario_3_4}). Notably, bias was constantly negligible, including for MAIC-1 and MAIC-2. Coverage was near nominal for anchored estimators, but unanchored PSW produced overly narrow confidence intervals, resulting in undercoverage.

Under DGM-4, which featured a lognormal distribution with positivity violation, estimator performance varied (Figure \ref{fig:treatment_scenario_3_4}). MAIC-1 and MAIC-2 yielded unbiased point estimates, while PSW retained some residual bias. Unanchored MAIC-1 demonstrated the highest precision. However, most estimators underestimated variability, leading to sub-nominal coverage rates for all estimators but unanchored MAIC-2 which overestimated variability, resulting in over-coverage.

\begin{figure}
    \centering
    \includegraphics[width=1\linewidth]{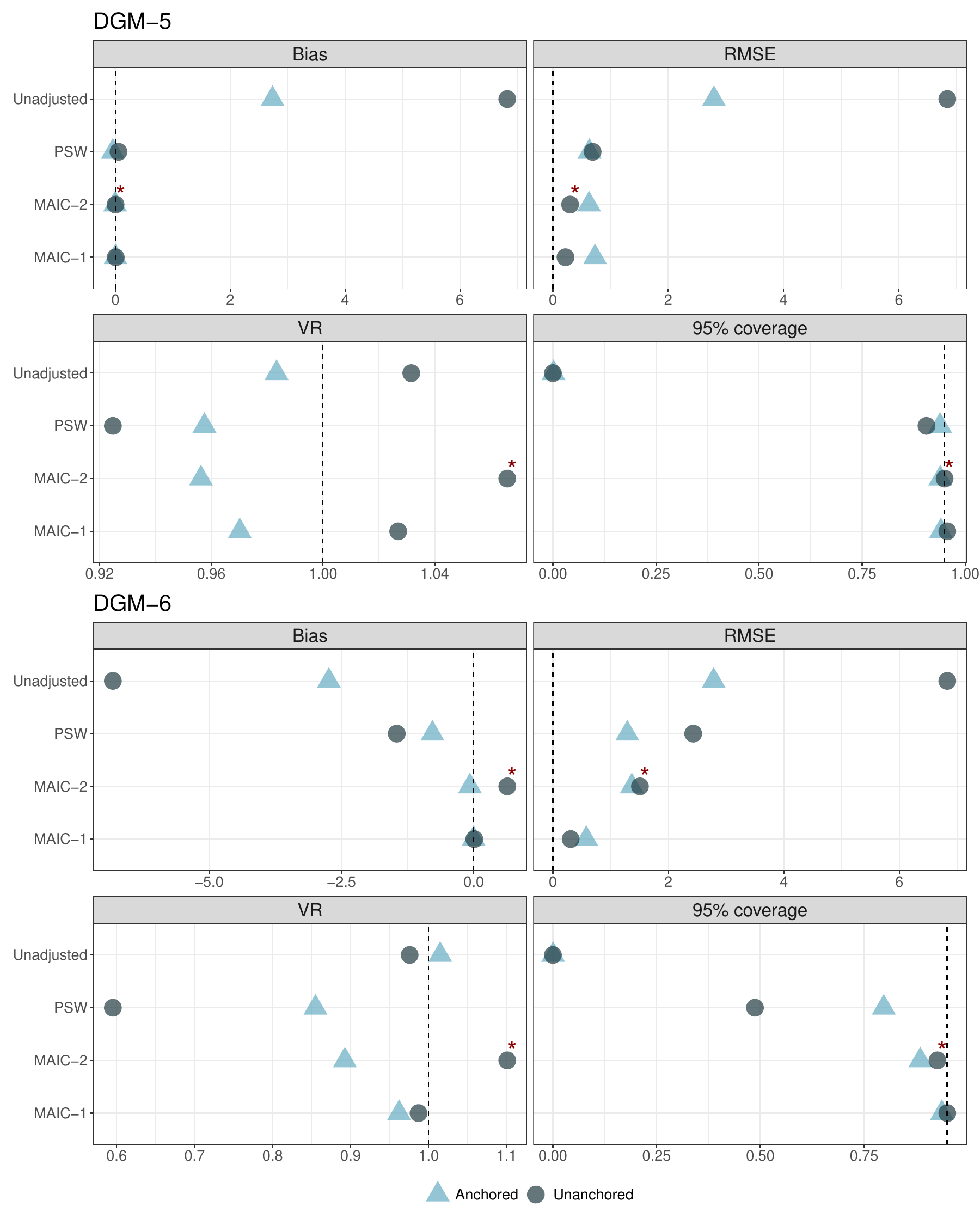}
    \caption{Treatment effect estimation under DGM-5 and DGM-6 (asterisks indicate that estimators did not converge for some of the iterations: DGM-6, 73/2000 for anchored MAIC-2, and 294/2000 for unanchored MAIC-2)}
    \label{fig:treatment_scenario_5_6}
\end{figure}

\begin{figure}
    \centering
    \includegraphics[width=1\linewidth]{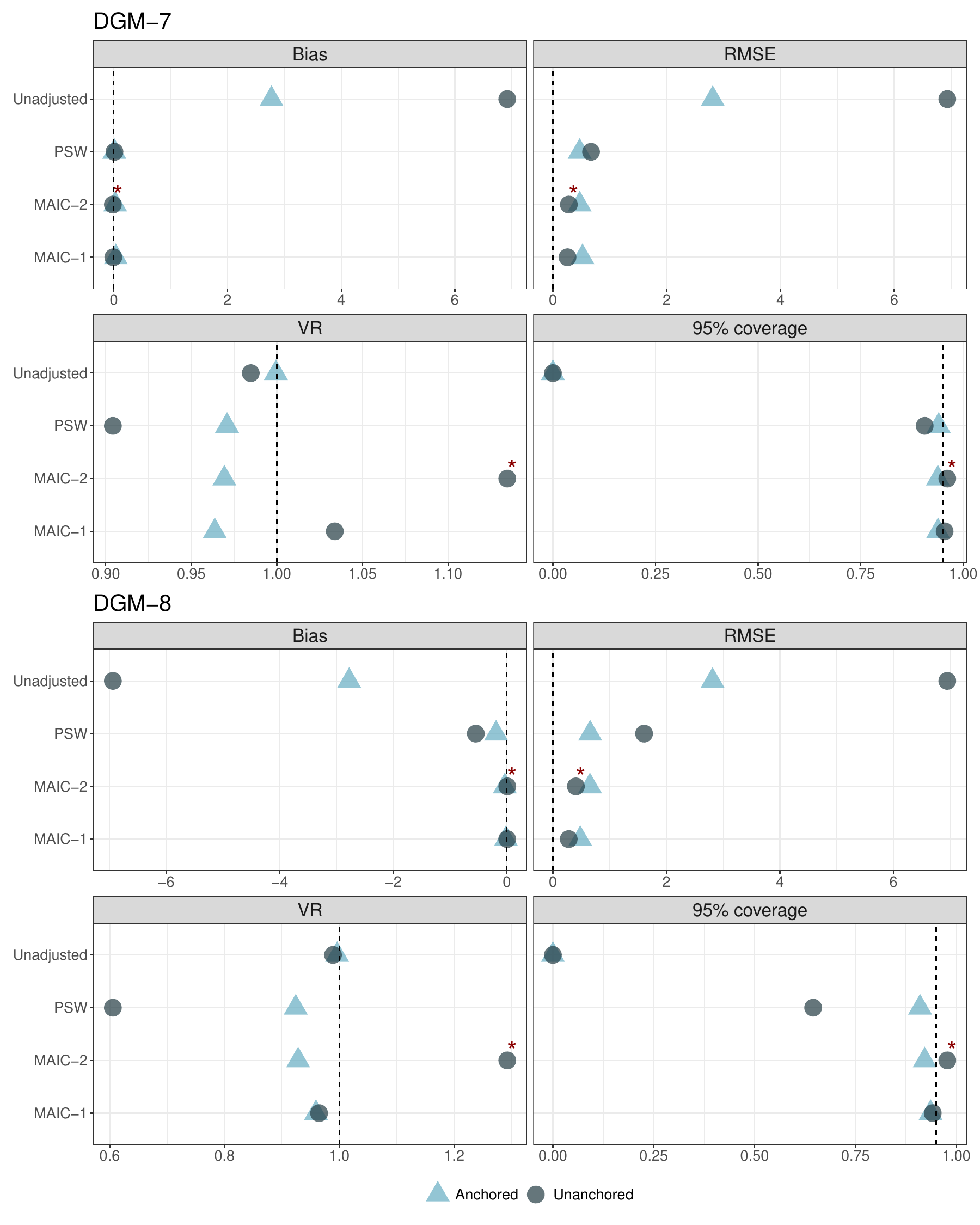}
    \caption{Treatment effect estimation under DGM-7 and DGM-8 (asterisks indicate that estimators did not converge for some of the iterations: DGM-7, 1/2000 for unanchored MAIC-2; DGM-8 4/2000 for unanchored MAIC-2)}
    \label{fig:treatment_scenario_7_8}
\end{figure}

Under DGM-5 and DGM-7 (with bimodal covariate distributions and no positivity violations), estimator performance was largely consistent with observations from DGM-1 (Figures \ref{fig:treatment_scenario_5_6} \ref{fig:treatment_scenario_7_8}). All estimators produced unbiased treatment effect estimates. Variability ratios remained close to unity across estimators, supporting well-calibrated standard errors and resulting in coverage rates near the nominal 95\% level. Notably, the bimodal nature of covariates did not meaningfully degrade the performance of the MAIC estimators, especially MAIC-1 which yielded the most precise treatment effect estimations in terms of bias and coverage rate.

Under DGM-6 and DGM-8, which combines bimodal covariate distributions with a lack of covariate overlap,  MAIC-1 and MAIC-2 showed the lowest bias and RMSE, outperforming PSW, which suffered from higher variability and bias (Figures \ref{fig:treatment_scenario_5_6} and \ref{fig:treatment_scenario_7_8}). In terms of coverage, only MAIC-1 (both anchored and unanchored) approached nominal levels without bias, while all other estimators demonstrated under or overcoverage. This suggests that MAIC-1 happens to be robust in scenarios with non-normal covariate distributions and non-trivial positivity violation, whereas weighting strategies that aim at balance the full covariate distribution are more sensitive under limited covariate overlap.

Additionally it is worth noting that in these scenario with bimodal covariate, the method of moments adjusting for first and second-order moments (MAIC-2) failed to converge properly or yielded absurd estimates on some instances due to excessively low effective sample sizes after weighting: DGM-5: 1/2000 (unanchored), DGM-6: 73/2000 (anchored) and 294/2000 (unanchored), DGM-7: 1/2000 (unanchored), and DGM-8: 4/2000 (unanchored).

\subsection{Comparisons between before and after weighting covariate distributions}

\begin{figure}
    \centering
    \includegraphics[width=1\linewidth]{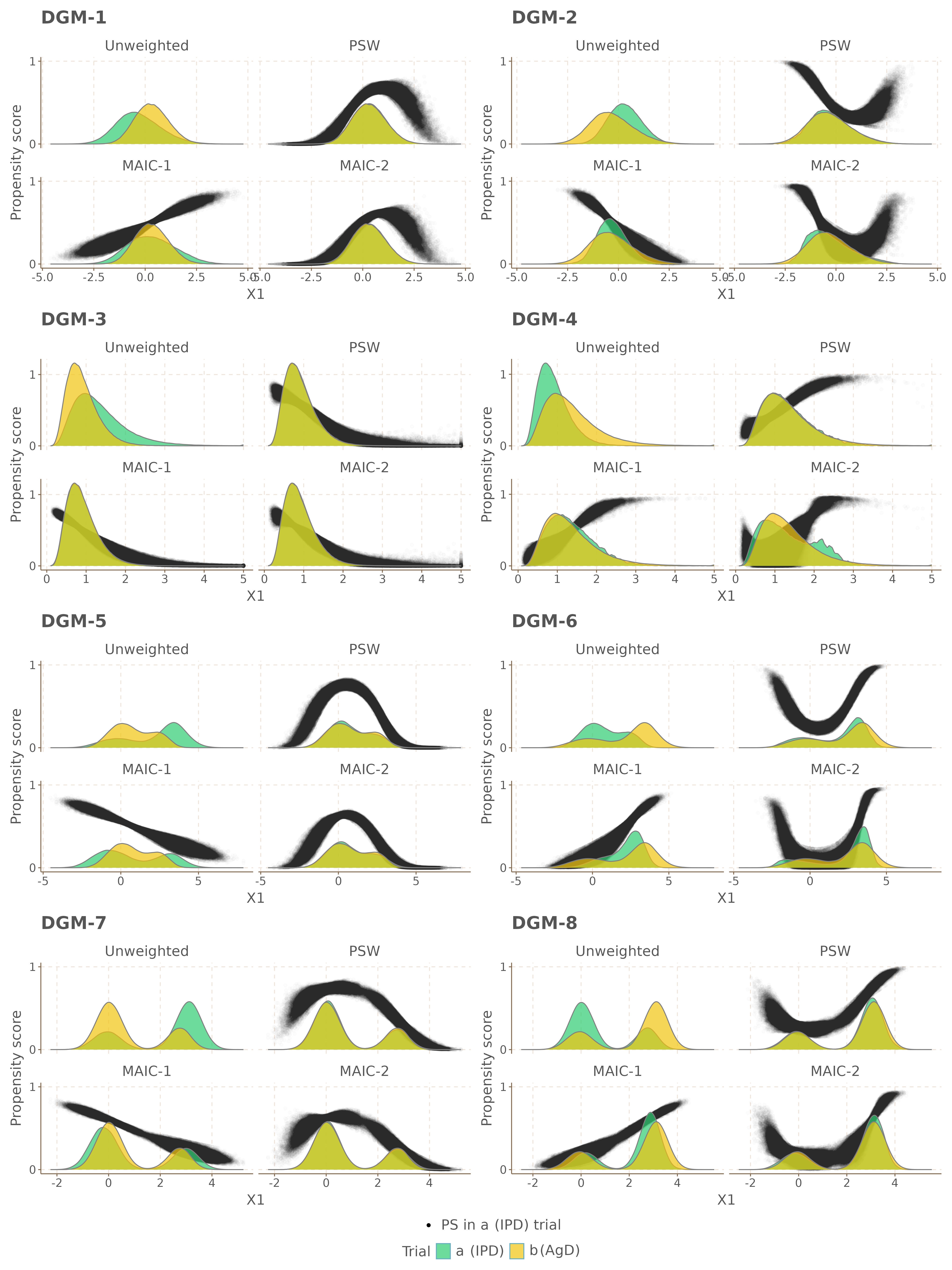}
    \caption{Anchored estimators: $X_1$ distribution before and after weighting + individual PS distribution depending on $X_1$ value}
    \label{fig:PS_anchored}
\end{figure}

An analysis of covariate distributions before and after weighting with anchored estimators revealed key differences across scenarios and estimators (Figure \ref{fig:PS_anchored}). Anchored results are presented in the main text because of the easier interpretation (adjusting for $X_1$ only). A corresponding figure showing weighting distributions for unanchored estimators is provided in Figure S1 \ref{fig:PS_unanchored}.
This figure displays the density distribution of the variable $X_1$ in trials $a$ and $b$, stratified by estimator. A second Y-axis represents the individual propensity score within the $a$ trial as a function of $X_1$. For each DGM, these plots aggregate observations coming from a sample of 100 iterations.

In DGM-1, all three weighting estimators effectively reweight the covariate distributions, though their post-weighting distributions differ: MAIC-2 and PSW produce distributions that closely match the target distribution trial $b$, while MAIC-1 only aligns the covariate means, consistent with its theoretical design. A similar pattern is observed in DGM-5 and DGM-7: PSW and MAIC-2 again replicate the $X_1$ distribution of trial $b$ post-weighting. Notably, MAIC-2 achieves this even with a bimodal distribution. As in DGM-1, the MAIC-1 post-weighting distribution of $X_1$ are not perfectly matching, but the means are equivalent.

In scenarios with positivity issue (\textit{i.e.} DGM-2, DGM-4, DGM-6, and DGM-8) limited statistical support for the target population in trial $a$ led to extreme propensity scores estimates under PSW, and to a lesser extent under MAIC-2 (\textit{i.e.}, propensity scores close to 1, resulting in very large weights), reflecting their intrinsic tendency to tightly match the target distribution. In contrast, MAIC-1 produces fewer extreme scores, yielding more stable weights. This observation may explain the results regarding the relatively less biased treatment effect estimates yielded by MAIC-1 as compared to PSW and MAIC-2.

\subsection{Sensitivity to incorrect adjustment model specification}

\begin{figure}
    \centering
    \includegraphics[width=1\linewidth]{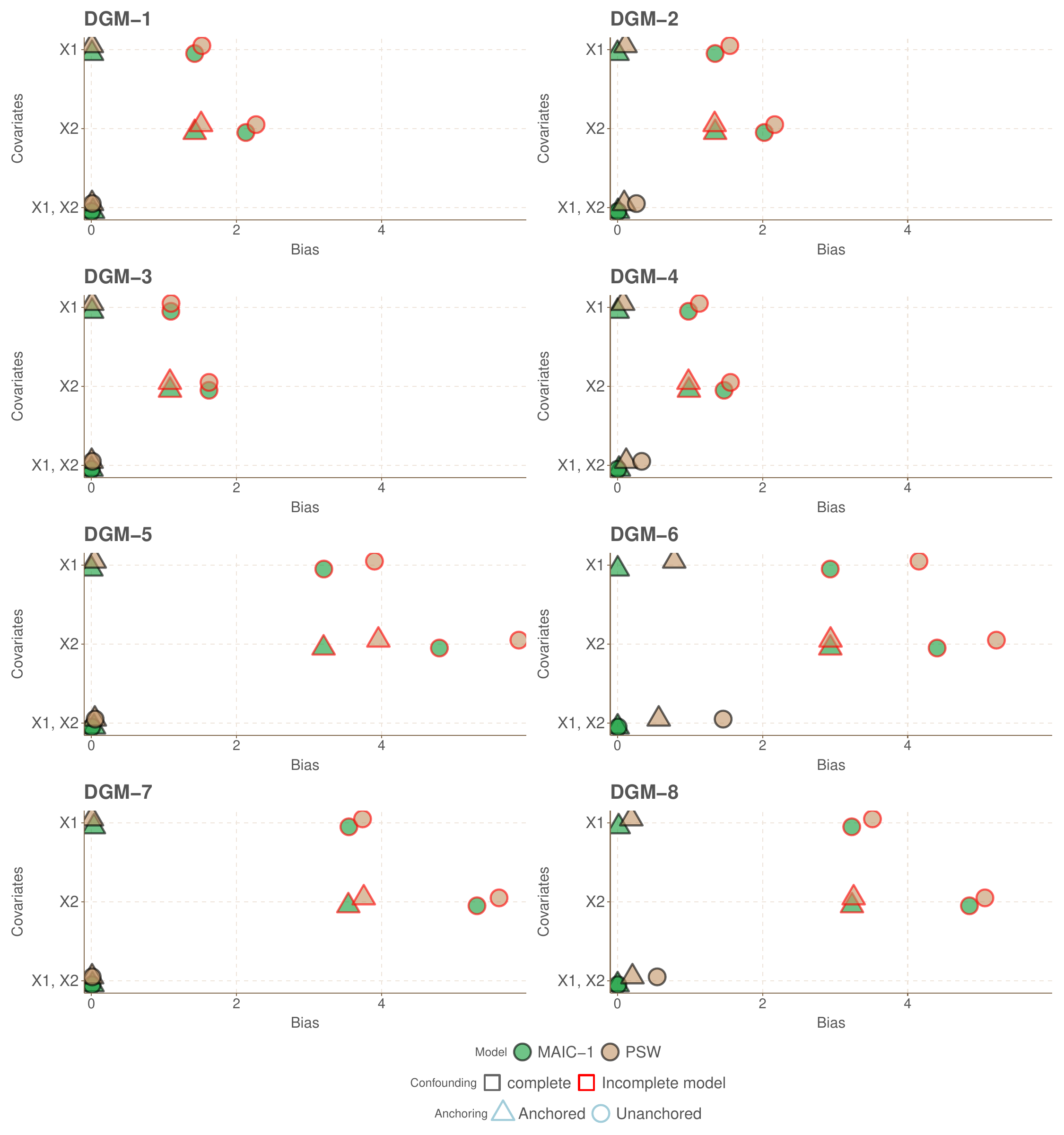}
    \caption{Bias of estimators depending on adequate model specification (the asterisk indicates that 3 iterations did not converge for unanchored MAIC-1)}
    \label{fig:bias_confounding}
\end{figure}

Figure~\ref{fig:bias_confounding} shows the bias of different estimators depending on whether treatment effect estimation models were adequately specified—that is, whether the trial assignment model included all relevant confounders. The Y-axis indicates the covariates included in the weighting model: either $X_1$, $X_2$, or both $X_1$ and $X_2$. In order to take into account all the relevant confounders, the anchored estimators should include $X_1$ (TEM), and the unanchored ones should include $X_1$ and $X_2$ (TEM and PF). Only MAIC-1 and PSW estimators are reported here in order to avoid overplotting, as the results for MAIC-2 were intermediate between these two estimators.

Bias due to model (mis-)specification was notably greater than that attributable to the choice of estimators. When key confounders were omitted, all estimators displayed substantial bias regardless of the weighting approach. Conversely, when all confounders were included, bias was markedly reduced across estimators. The only substantial difference in bias across different weighting methods with correct model specification was observed in the context of DGM-6 and DGM-8.

\subsection{Empirical application}

\subsubsection{The AKIKI and AKIKI-2 randomized controlled trials}

\begin{figure}
    \centering
    \includegraphics[width=1\linewidth]{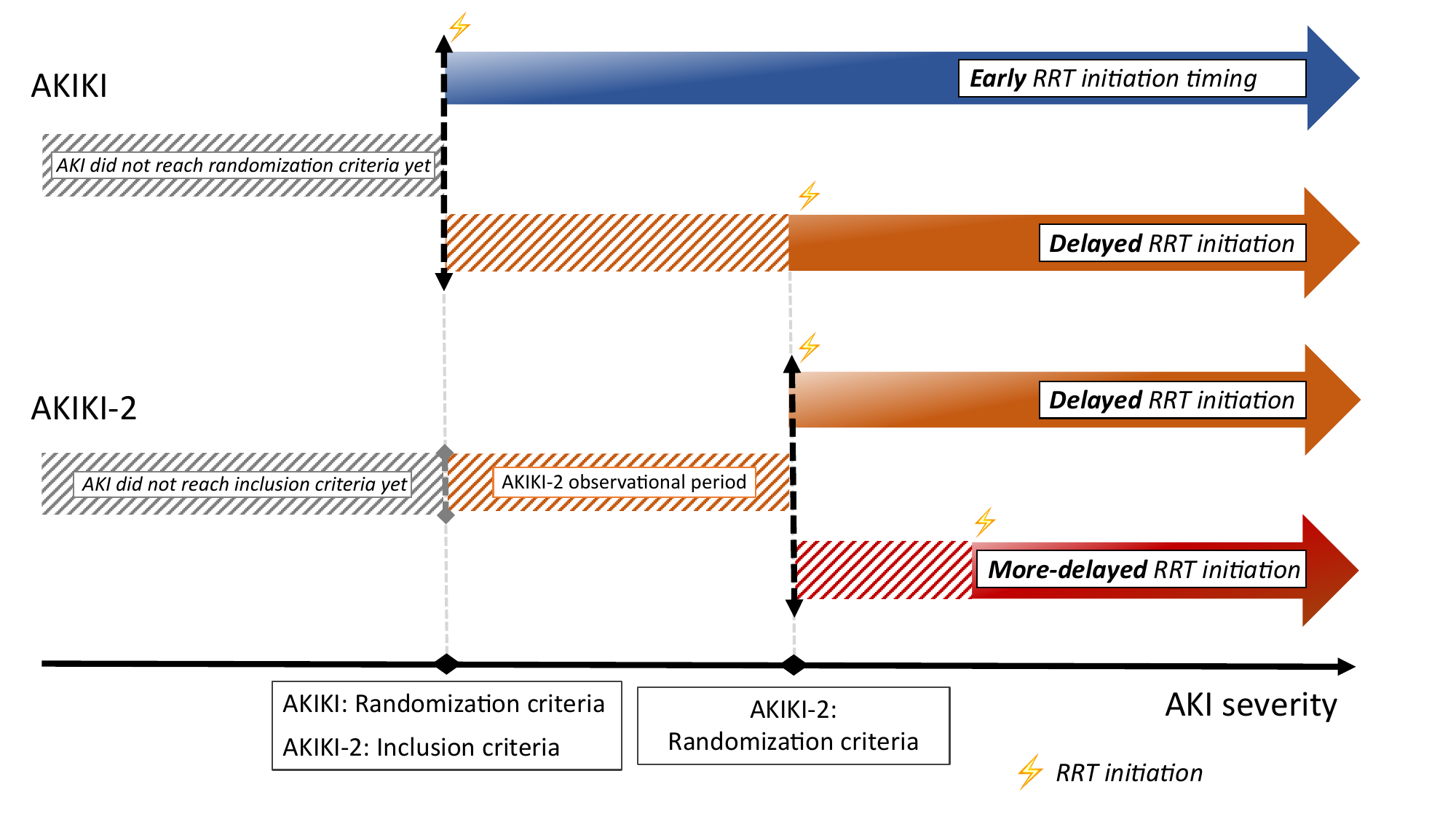}
    \caption{AKIKI trials: treatment arms according to the RRT initiation time}
    \label{fig:schema_akiki}
\end{figure}

We conducted an empirical evaluation of the different estimators tested in the Monte Carlo simulations using data from the AKIKI and AKIKI-2 RCTs. Both trials investigated strategies for initiating renal replacement therapy (RRT) in critically ill patients with acute kidney injury, differing primarily in the timing of initiation.
The AKIKI trial compared an "early" RRT initiation strategy, where RRT was started upon meeting predefined inclusion criteria, to a "delayed" strategy that deferred initiation until more severe conditions arose. AKIKI-2 extended this comparison by evaluating the same "delayed" strategy from AKIKI against a "more-delayed" approach further relaxing RRT initiation criteria. Inclusion criteria in the AKIKI-2 trial were identical to AKIKI's; however after inclusion patients first entered an observational phase until they eventually met the "delayed" strategy RRT initiation criteria, at which time they were randomized into either "delayed" or "more-delayed" RRT treatment arms. Figure \ref{fig:schema_akiki} illustrates these three treatment arms evaluated across the two AKIKI trials.

Both trials found no statistically significant differences in survival at 60 days. The presence of a common comparator arm ("delayed" strategy), and comparable inclusion criteria made these trials well-suited for evaluating different indirect comparisons strategies.

To assess estimators, we alternatively considered AKIKI and AKIKI-2 as the target population. In alignment with the simulation settings, we evaluated both anchored and unanchored strategies, as well as estimators fitted with different sets of variables: prognostic factors (Age, Minimum Mean Arterial Pressure, SOFA score), treatment effect modifiers (Chronic Kidney Disease, SpO2/FiO2 ratio, 24h diuresis), or both. These variables sets were selected based on the clinical expertise of AKIKI trials investigators guidance and data analysis. Additionally, we considered two approaches for aligning the trials depending on the reference starting date considered in the AKIKI-2 trial: the date of start of follow-up (observational period), or the date of randomization. Using the date of inclusion (start of follow-up) is the most appropriate alignment since inclusion criteria were identical at that point in both AKIKI and AKIKI-2. However, if relying on published data, the date of randomization would likely be used instead. To ensure comparability of AKI severity at baseline across trials when aligning on AKIKI-2 inclusion date, we randomly assigned in a 1:1 ratio patients from AKIKI-2 observational data who were eligible for AKIKI but never reached AKIKI-2 randomization criteria. This emulated a hypothetical allocation to "delayed" vs. "more-delayed" RRT; for these patients who never reached the "delayed" (and thus "more-delayed") RRT initiation threshold, the observed outcomes correspond to those that would have been observed under randomization.

Crucially, the choice of alignment date has a major impact on the covariate distribution across trials. Aligning on the date of inclusion yields broadly comparable populations, whereas aligning on the date of randomization results in substantial differences in baseline characteristics, leading to positivity issues in some instances, thus reflecting the range of scenarios explored in our simulations.

\subsubsection{Illustration of estimator behavior using AKIKI data}

\begin{figure}
    \centering
    \includegraphics[width=1\linewidth]{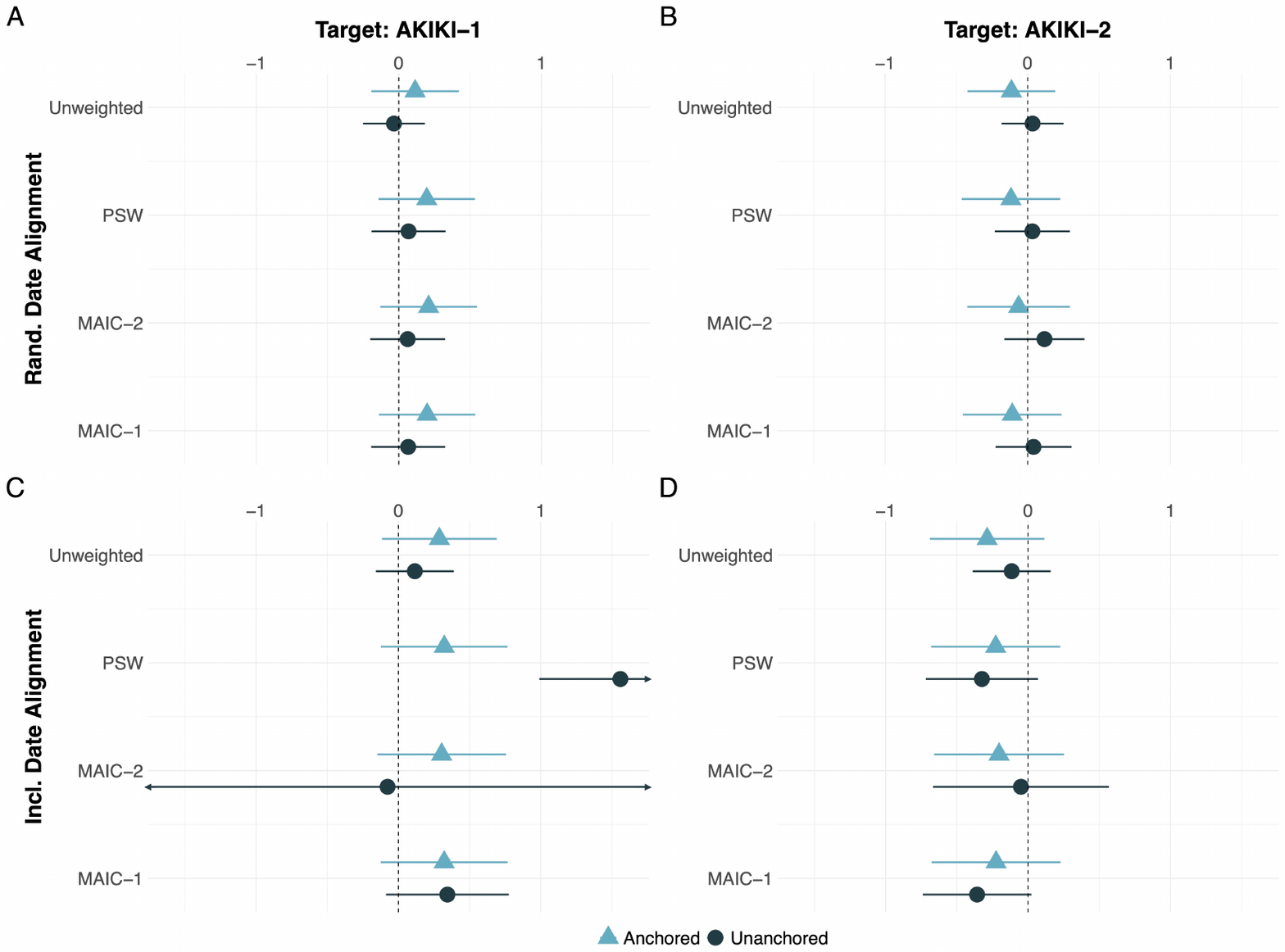}
    \caption{Empirical data application on AKIKI RCTs: log(HR) estimate (95\% CI) for the "more-delayed" strategy vs "early" strategy, using the "delayed" strategy as the anchor. The four panels differ on the target population (AKIKI vs AKIKI-2), and the alignment date according to the AKIKI-2 inclusion or randomization date. 95\% Confidence intervals are not displayed on panel C for unanchored PSW and MAIC-2 because bounds failed outside of the plot limits.}
    \label{fig:akiki_results}
\end{figure}

Figure \ref{fig:akiki_results} presents the results of the empirical evaluation of the different estimators using data from the AKIKI and AKIKI-2 clinical trials. The four panels differ by the target population and therefore the IPD used (panel A and C: target population is AKIKI and IPD is AKIKI-2, and panels B and D: target population is AKIKI-2 and IPD is AKIKI). These findings can be summarized as follows. (i) The variability across estimators within the same anchoring approach (anchored or unanchored) is considerably smaller than the variability observed between anchored and unanchored approaches. (ii) When unanchored comparisons are used and the adjustment set is expanded to include prognostic factors, results exhibit greater variability and become more sensitive to the alignment date.

Specifically, when AKIKI-2 is used as the target population (panels B and D), estimates are relatively consistent across estimators and alignment strategies, regardless of the confounding adjustment strategy. In contrast, when AKIKI is used as the target population and the date of randomization of AKIKI-2 is used as the follow-up starting date (Panel C), unanchored estimators exhibit markedly wider confidence intervals, particularly for MAIC-2 and PSW. This reflects a substantial reduction in effective sample size due to poorer covariate overlap, with MAIC-1 maintaining more stable estimates and narrower confidence intervals under these conditions. While we know from theory that this alignment date is incorrect, it may have been the one used in practical settings if basing the analysis on reported results of the RCT, which stresses out the sensitivity of these methods to the lack of statistical support.

\section{Discussion}

This study evaluated the performance of MAIC estimators in comparison to a PSW estimator that leverages full individual-level data, using Monte Carlo simulations across diverse scenarios. The goal was to examine how deviations from ideal conditions---such as violations of the positivity (overlap) assumption or non-normal covariate distributions---affect estimators' performance. Several key findings emerged from our investigation. First, under ideal conditions with normally distributed covariates, sufficient overlap between trials, and correctly specified adjustment models, all population-adjusted estimators produced unbiased treatment effect estimates with coverage near the nominal 95\% level. In this setting, the unanchored approach showed slightly better precision (lower RMSE) than anchored analysis, consistent with the intuition that using a common comparator involves an additional level of variability. Similarly, when introducing non-normal covariate distributions without overlap issues, all estimators maintained unbiasedness and nominal coverage, suggesting that the MAIC method of moments weighting remains relevant even with non-normal covariate distributions. This result offers some empirical reassurance regarding a previously noted concern that the theoretical properties of MAIC under non-standard covariate distributions were not well understood.

In contrast, when the DGM involved positivity violations, clear differences between estimators emerged, and the full-IPD PSW estimator exhibited persistent bias despite correct model specification. This indicates that even with all relevant covariates included, the IPD-based estimators struggled to extrapolate beyond the support of the data, consistently with known properties of PSW \citep{zhuCoreConceptsPharmacoepidemiology2021} \citep{liAddressingExtremePropensity2019}. The MAIC estimators, particularly the simpler variant adjusting only for the first moment (MAIC-1), by contrast, remained less biased and achieved acceptable coverage, suggesting that method of moments weighting can be more robust to moderate overlap issues. However, it must be noted that MAIC approaches encountered convergence issues in a small number of simulation iterations likely due iterations with more extreme overlap issues.

We also investigated the impact of model misspecification on the estimators by deliberately omitting a confounder from the adjustment model. The results of this analysis demonstrated that bias was driven primarily by whether the model correctly included all treatment effect modifiers, rather than by the choice of weighting method, consistently with literature results \citep{hatswellEffectsModelMisspecification2020} \citep{campbellEstimatingMarginalTreatment2023}.


Our work builds upon and extends the existing literature on population-adjusted indirect comparisons. Previous applied studies had compared MAIC with estimators using only IPD, but mostly in the context of specific case studies without extensive simulation. For example, Park et al. used a case study in small-cell lung cancer to compare various population-adjustment estimators (MAIC, simulated treatment comparison, and doubly robust combination of the two) and estimators using full IPD (propensity weighting, outcome regression, and doubly robust)\citep{parkUnanchoredPopulationAdjustedIndirect2024}. They reported that the resulting treatment effect estimates were fairly similar across methods, which was attributed to the fact that the two trial populations in their case study were not significantly different. Similarly, Wong et al.
 compared treatment outcomes for vedolizumab vs. adalimumab in ulcerative colitis using either a PSW or an MAIC approach, and found only minimal differences between the adjusted estimates from these two methods \citep{wongMatchingadjustedIndirectComparisons2023}. These empirical comparisons align with the portion of our findings showing that when populations have substantial overlap, MAIC and PSW yield concordant results.

However, prior to our work, there had not been a comprehensive simulation study systematically stress-testing MAIC against an IPD estimator across a range of less ideal scenarios. A recent study by Campbell et al. conducted two separate simulation exercises to compare weighting vs. outcome regression estimators, under settings with and without full-IPD availability, but they did not directly cross-compare aggregated data and full-IPD approaches\citep{campbellEstimatingMarginalTreatment2023}.

The results confirm some of the theoretical expectations discussed in earlier methodological literature. For instance, methodological reviews by Phillippo et al., and Remiro-Azócar et al. have cautioned that unanchored indirect comparisons rely on strong assumptions (no unmeasured confounding) and may hardly be trustworthy in real-life applications \citep{phillippoMethodsPopulationAdjustedIndirect2018} \citep{remiro-azocarMethodsPopulationAdjustment2021}. While not new, we provide additional evidence for these cautions. Our results emphasize that whether IPD are fully accessible is of a much lower importance as compared to having a common treatment arm to form the indirect comparison.

A strength of this work is the various data-generating mechanisms considered, namely comparing distribution shapes (normal vs. bimodal), while varying degree of population overlap. We also augmented the simulation study with a real-world example, with consistent results. To our knowledge, this is the first study to  compare using Monte Carlo simulations a population-adjusted indirect comparison technique against its full-IPD counterpart. Additionally, by including both anchored and unanchored implementations, and comparing complete vs. biased models, we shed light on the relative importance of these different parameters in the eventual performance of the estimators.

Several limitations of our study should be acknowledged. First, our simulations focused on a continuous outcome and a linear model for the treatment effect. Most real-world applications involve non-collapsible effect measures (e.g. odds ratios for binary outcomes or hazard ratios for time-to-event outcomes). The relative performance of MAIC and PS weighting in those contexts remains to be investigated. Second, the finding that MAIC outperformed the full-data PSW in scenarios with overlap violations (DGMs 2, 4, 6 and 8) must be interpreted in context. In practice, if an analyst has actually access to IPD from both trials, other estimands can be targeted which may be less subject to positivity issue depending on the context: Average Treatment Effect in the Treated, or Average Treatment Effect in the Overlap population \citep{liAddressingExtremePropensity2019}. Closely related to that aspect, some remedial measures could be taken, such as truncating portions of the sample distributions that lie outside the common support. By doing so, one can largely resolve positivity violations and may obtain more reliable estimates with IPD methods. However, this procedure also changes the target population: the estimand is no longer the treatment effect in the original target trial’s population, but rather in the overlapped sub-population that both trials share. Whether this modified estimand is of interest depends on the context. In our simulations, we did not implement trimming for the PSW estimator; we evaluated the method on the original target estimand even when overlap was poor, to observe the unmitigated impact of positivity violations. In a real analysis with full IPD, one would not usually proceed with a gross overlap violation without some corrective action. Therefore, situations where MAIC appeared to outperform PSW (in bias or coverage) may reflect settings where corrective measures could be applied in practice, potentially mitigating PSW's shortcomings. In other words, our results do not imply that an aggregate-data approach is superior to an IPD approach in absolute terms, but rather that the method of moments actually yields satisfactory results in some situations with overlap issues. As a consequence, in situations where the estimand of interest is actually the one targeted by MAIC and when one assumes positivity issues, matching the first moment of the distribution may be of interest, regardless of the actual availability of individual data. Whether that estimand is actually the primary estimand of interest, particularly when compared with the Average Treatment Effect in the combined trials population, has been discussed elsewhere \citep{remiro-azocarTargetEstimandsPopulationadjusted2022, russek-cohenDiscussionTargetEstimands2022}.

We did not evaluate other approaches such as Simulated Treatment Comparison (STC) or meta-regression methods that incorporate both IPD and aggregate data, nor more complex extensions like multilevel network meta-regression. These methods could be examined in future work to see if they share similar properties or offer advantages in the scenarios we considered.

Finally, as with any simulation study, our scenarios are simplification of real-world situations. We assumed two covariates and a rather simple outcome model structure; actual studies may involve multiple effect modifiers or non-linear effects. Moreover, other non-normal covariate distributions could prove challenging for an MAIC estimator, especially a covariate distribution which moments may not be defined.


Despite these limitations, our findings have important implications for the conduct and interpretation of indirect treatment comparisons. They provide concrete evidence that when all assumptions are met, population-adjusted estimators like MAIC can perform just as well as conventional PSW that uses complete data, even in some cases maintaining accuracy in situations where the latter suffers (e.g. moderate overlap violations). This is an encouraging result when one must rely on aggregate data out of necessity since it suggests that, under reasonably good conditions, using MAIC does not inherently compromise the validity or precision of the estimated treatment effect. Consequently, the added value of access to full IPD lies in greater analytical flexibility. At the same time, our results emphasize that the primary determinants of validity are the upholding of causal inference assumptions---positivity, correct model specification and exchangeability---thus stressing out once again the inherent superiority of anchored as compared to unanchored comparisons.

\newpage

\paragraph{Author contributions}
ASL, DH, JL conceived the study and developed the statistical methodology of the simulation study. ASL, DH, JL and SG conceived the empirical application. ASL implemented the simulations and empirical application. DH supervised the statistical analyses. ASL, DH, JL and SG interpreted the results. ASL drafted the first version of the manuscript. DH, JL and SG critically revised the manuscript. All authors read and approved the final version of the manuscript.

\paragraph{Financial disclosure}

None reported.

\paragraph{Conflict of interest}

\noindent {ASL: Remunerated lecture for Pierre Fabre Laboratories on PAICs, unrelated to this work}

\listoffigures
\processdelayedfloats

\bibliography{My_Library}

\clearpage
\appendix
\section{Appendix}
\renewcommand{\thefigure}{S\arabic{figure}}
\renewcommand{\thepostfigure}{S\arabic{postfigure}}
\setcounter{figure}{0}
\setcounter{postfigure}{0}

\begin{figure}
    \centering
    \includegraphics[width=1\linewidth]{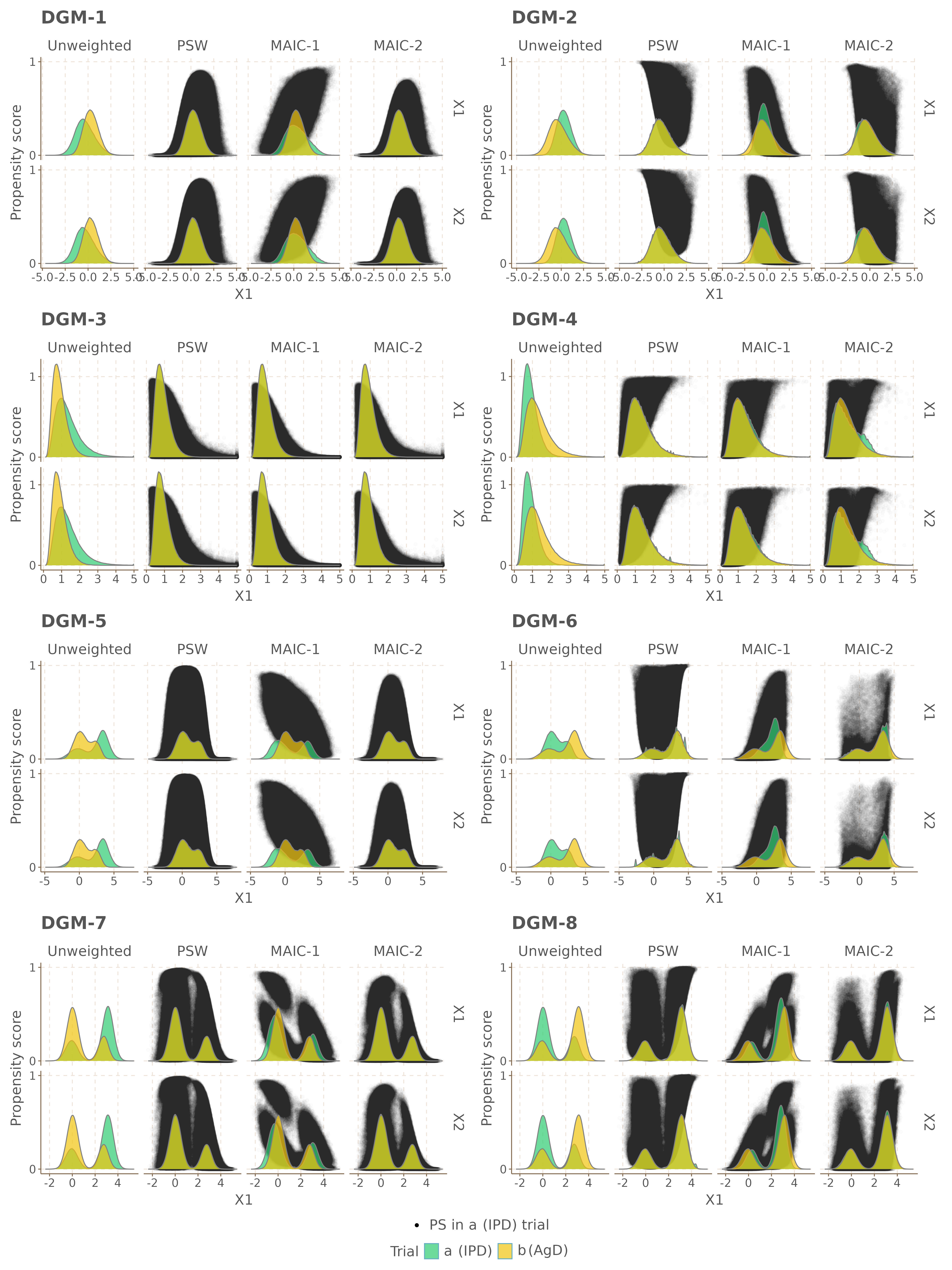}
    \caption{Unanchored estimators: $X_1$ and $X_2$ distributions before and after weighting + individual PS distribution depending on $X_1$ and $X_2$ values}
    \label{fig:PS_unanchored}
\end{figure}
\processdelayedfloats

\end{document}